\documentclass[final,5p,times,twocolumn]{elsarticle}

\usepackage{amssymb}

\journal{Computer Science Review}

\usepackage{array}
\usepackage{graphicx}
\usepackage{textcomp}
\usepackage{multirow}
\usepackage{booktabs}
\usepackage[caption=false]{subfig}
\usepackage[table]{xcolor}
\usepackage{easyReview}
\setreviewsoff

\begin{document}

\begin{frontmatter}

\title{Asynchronous Federated Learning on Heterogeneous Devices: A Survey}

\author[aff1]{Chenhao Xu}
\address[aff1]{
Deakin Blockchain Innovation Lab, School of Information Technology, Deakin University, Geelong, VIC, Australia
}

\author[aff2]{Youyang Qu}
\address[aff2]{
Shandong Computer Science Center; Qilu University of Technology, China
}

\author[aff1]{Yong Xiang}

\author[aff2]{Longxiang Gao}

\fntext[aff1]{Youyang Qu and Longxiang Gao are the corresponding authors.}

\begin{abstract}
\add{Federated learning (FL) is a kind of distributed machine learning framework, where the global model is generated on the centralized aggregation server based on the parameters of local models, addressing concerns about privacy leakage caused by the collection of local training data. With the growing computational and communication capacities of edge and IoT devices, applying FL on heterogeneous devices to train machine learning models is becoming a prevailing trend. Nonetheless, the synchronous aggregation strategy in the classic FL paradigm, particularly on heterogeneous devices, encounters limitations in resource utilization due to the need to wait for slow devices before aggregation in each training round. Furthermore, the uneven distribution of data across devices (i.e. data heterogeneity) in real-world scenarios adversely impacts the accuracy of the global model. Consequently, many asynchronous FL (AFL) approaches have been introduced across various application contexts to enhance efficiency, performance, privacy, and security. This survey comprehensively analyzes and summarizes existing AFL variations using a novel classification scheme, including device heterogeneity, data heterogeneity, privacy, and security on heterogeneous devices, as well as applications on heterogeneous devices. Finally, this survey reveals rising challenges and presents potentially promising research directions in this under-investigated domain.}
\end{abstract}




\end{frontmatter}

\section{Introduction}
\label{sec:introduction}

\add{In recent years, the rapid expansion of computational capabilities coupled with the swift evolution of communication infrastructures has directly contributed to the flourishing of machine learning (ML), a pivotal driving force behind numerous contemporary technologies~\cite{ghahramani2015probabilistic, xu2020reinforcement}. Nonetheless, the training of ML models necessitates a substantial volume of high-quality data, a crucial requirement for model trainers operating in real-world scenarios~\cite{nie2020semi}. Notably, the emphasis on preserving privacy during data sharing continues to persist, with newly enacted legislation and regulations further complicating the process of data acquisition~\cite{voigt2017eu, andrew2021general, wachter2019right}. Furthermore, industries exhibit hesitancy in sharing their local data due to competitive pressures, privacy concerns, and other potential considerations~\cite{qu2018privacy, xiong2018attribute, xiong2020edge}. All these factors jointly give rise to the challenge of \textit{isolated data islands}. As a result, gathering data from diverse reliable sources becomes a near-impossible task, often accompanied by prohibitively high costs~\cite{xie2018data, su2020lvbs}.}

\add{Federated learning (FL) presents an innovative framework facilitating collaborative ML model training among multiple entities without requiring direct access to their respective local training data. Initially introduced by Google in 2016, FL emerges as a promising ML approach that effectively addresses the imperatives of data privacy and communication efficiency~\cite{konevcny2016federated1, konevcny2016federated2}. The fundamental objective of FL revolves around ensuring personal data privacy and engendering robust ML models across multiple participants or computational nodes while upholding legal mandates~\cite{lim2020federated, yin2021comprehensive}. Consequently, FL has found its way into numerous research papers, employing a central server to collect parameters of local models from nodes (referred to as ``local models'' henceforth) prior to aggregating them into a global model during each round of training~\cite{wahab2021federated}.}

\add{Amidst the extensive rollout of the 5G network and the swift evolution of hardware capabilities, heterogeneous devices, encompassing both edge and IoT devices, are experiencing augmented communication and computational capabilities, paving the way for an expanded array of applications~\cite{imteaj2021survey}. Compared with classic ML approaches, FL presents a range of merits specifically tailored to edge applications~\cite{lim2020federated, khan2021federated}: (1) Enhanced preservation of local data privacy, facilitated by the gradients-based aggregation of the global model; (2) Reduced network transmission latency, as the training data remains localized instead of being transmitted to cloud servers; (3) Elevated model quality, owing to the incorporation of learned features from other devices. Consequently, FL serves as a catalyst for collaborative ML model training across heterogeneous devices, a phenomenon well-documented in numerous research publications.}

\begin{table*}[ht]
    \footnotesize
    \renewcommand{\arraystretch}{1.2}
    \caption{Comparison with Existing Surveys}
    \label{table:survey-compare}
    \centering
    \begin{tabular}{c|c|l}
    \toprule
    \textbf{Surveys} & \textbf{Topics} & \textbf{Limitations} \\
    \midrule
    \cite{wahab2021federated} 
    & FL & Multi-level classification of FL without a detailed classification of AFL. \\
    \hline
    \cite{lim2020federated} 
    & FL on edge & Treat AFL as a promising solution without comparing different AFL schemes. \\
    \hline
    \cite{imteaj2021survey} 
    & FL on IoT & Only 7 papers related to AFL are gathered with an investigation on the convergence. No detailed classification on AFL. \\
    \hline
    \cite{khan2021federated} 
    & FL on IoT & Explain the concept of AFL without comparing different AFL schemes. \\
    \hline
    \cite{yin2021comprehensive} 
    & FL privacy & Focus on privacy-preserving in FL, no discussion related to AFL. \\
    \hline
    \cite{lyu2020threats} 
    & FL security & AFL not mentioned. \\
    \hline
    \cite{abdel2022federated} 
    & FL privacy & Focus on privacy-preserving in FL on IoT, no detailed classification of AFL. \\
    \hline
    This Survey & AFL & Classify and analyze the challenges faced by AFL and summarize the application scenarios of AFL. \\
    \bottomrule
    \end{tabular}
\end{table*}

\add{When employing classical FL on devices with limited resources, several drawbacks become apparent~\cite{wu2020safa}: (1) Device Unreliability. The presence of heterogeneous devices introduces a challenge as the aggregation server must wait for updated local gradients from chosen heterogeneous devices. These devices, however, might unexpectedly go offline due to their inherent unreliability. (2) Aggregation Efficiency Reduction. In each training round, faster devices are forced to wait for stale local models from slower devices (stragglers). This delay results from the dual factors of device heterogeneity (resource variation among devices) and data heterogeneity (uneven training data distribution across devices). (3) Low Resource Utilization. The current inefficiencies in node selection algorithms often cause multiple competent devices to be rarely chosen for participation. (4) Security and Privacy Vulnerabilities. The classic FL approach is susceptible to various security threats, such as poisoning and backdoor attacks. Moreover, concerns about privacy arise due to potential data leaks during the training process.}

\add{To address the challenges of device unreliability, aggregation efficiency reduction, and low resource utilization, asynchronous federated learning (AFL) emerges as a promising solution. In AFL, the central server promptly initiates global model aggregation upon the reception of a local model. As the devices unexpectedly going offline are ignorable to AFL, the concerns about device unreliability are mitigated. By removing the necessity to await slow devices for local model uploads before aggregation, AFL enhances aggregation efficiency. AFL also improves the utilization of computing resources across heterogeneous devices by allowing devices with varying operational efficiency to train their respective local models at their own pace.}

\add{While there have been survey papers on the subject of FL, none of them have undertaken an exhaustive investigation, classification, or summary of AFL. Consequently, the primary significance of this study lies in its comprehensive classification, summarization, and analysis of AFL. A comparative overview between this survey paper and other relevant surveys is provided in Table~\ref{table:survey-compare}.}

\add{The contributions of this survey paper are summarized as follows. Firstly, this comprehensive survey reviews and analyzes 125 research papers spanning the years 2019 to 2022, including 7 relative survey papers. Secondly, the existing papers of AFL are categorized and summarized innovatively from the perspective of device heterogeneity, data heterogeneity, privacy and security, and application scenarios. Thirdly, the survey identifies a number of promising research topics that deserve more investigation and discussion.}

The subsequent sections of this paper are organized as follows: In Section~\ref{sec:background}, the preliminary knowledge requisite for this survey is briefly introduced. Then, the AFL approaches addressing various challenges, including device heterogeneity, data heterogeneity, as well as privacy and security on heterogeneous devices, are then summarized and analyzed in Section~\ref{sec:device-heterogeneity}, Section~\ref{sec:data-heterogeneity}, and Section~\ref{sec:privacy-security}, respectively. Following this, Section~\ref{sec:applications} offers a comprehensive portrayal of the diverse applications of AFL with heterogeneous devices. Building upon the comprehensive analysis and discourse, Section~\ref{sec:research-directions} outlines potential avenues for promising research directions, followed by a conclusion in Section~\ref{sec:conclusion}.

\section{Background Knowledge}
\label{sec:background}

This section provides an explanation of the foundational knowledge essential for this survey, covering three key perspectives: federated learning, blockchain, and differential privacy.

\subsection{Federated Learning}
\label{sec:bg_fl}

\add{Distributed ML (DML) is a research topic that investigates different structures or topologies of the computer cluster for better training machine learning models~\cite{verbraeken2020survey}. Typically, DML can be categorized as centralized, decentralized, and fully distributed, along with various types of communication protocols. FL is a kind of DML where local models are trained on distributed nodes, and the global model is generated on an aggregation server by averaging the local models~\cite{mcmahan2017communication, liu2022distributed}. Nevertheless, there is a difference between FL and DML. FL is designed for scenarios where data resides on multiple devices or servers and is not centralized. It is typical in environments where data privacy is crucial, such as on mobile devices. By contrast, DML often assumes cloud or data center environments. While data can be distributed across multiple servers or nodes, these nodes are typically co-located or part of a single infrastructure, and there is more flexibility in data sharing. Therefore, FL serves as an ML framework designed to dismantle the barrier of the data silo~\cite{xu2023scei}, primarily attributed to its privacy-preserving characteristic for local training data within each node. The conventional FL process encompasses the following key steps.} 
\begin{enumerate}
\item Initialization: \add{Once the training task is defined according to the specific application scenario, the aggregation server prepares the initial global model $w_G^0$, alongside training parameters like learning rate, batch size, and iteration count.} Following the selection process, the aggregation server broadcasts $w_G^0$ to a designated number of nodes (denoted as $K$).
\item Local Model Training: Assume $t$ stands for the current iteration number. Based on global model $w_G^t$, each node trains its respective local model $w_k^t$, where $k \in [1, K]$. These local models, $w_k^t$, are subsequently transmitted back to the aggregation server.
\item Global Model Aggregation: \add{Assuming the training samples on node $k$ amount to $n_k$, and the total of training samples is represented as $n$, the server generates a new global model by taking a weighted average of the local models, as given in Eq.~\ref{eq:fedavg}.
\begin{equation}
w_G^{t+1} = \sum^{K}_{k=1} \frac{n_k}{n} w^t_k
\label{eq:fedavg}
\end{equation}
Subsequently, $w_G^{t+1}$ is sent back to the nodes in preparation for the next training iteration.}
\end{enumerate}

Generally, datasets across nodes in FL are expected to be independent and identically distributed (IID). \add{This entails an identical distribution of samples among nodes, with each training round independently selecting training samples. In practice, however, the training data samples collected by nodes usually deviate from this IID condition, named as non-independent and identically distributed (non-IID), posing challenges for both classic FL and AFL. Taking the smart hospital scenario as an example, IID data implies that disease cases across various hospitals exhibit similarity. Non-IID data, on the other hand, implies the diversity of disease cases among different hospitals, a portrayal more aligned with real-world scenarios. Under non-IID conditions, gradients learned and updated in one hospital provide limited utility in predicting patient conditions within another hospital.}

\begin{figure*}[ht]
    \centering
    \includegraphics[width=0.8\linewidth]{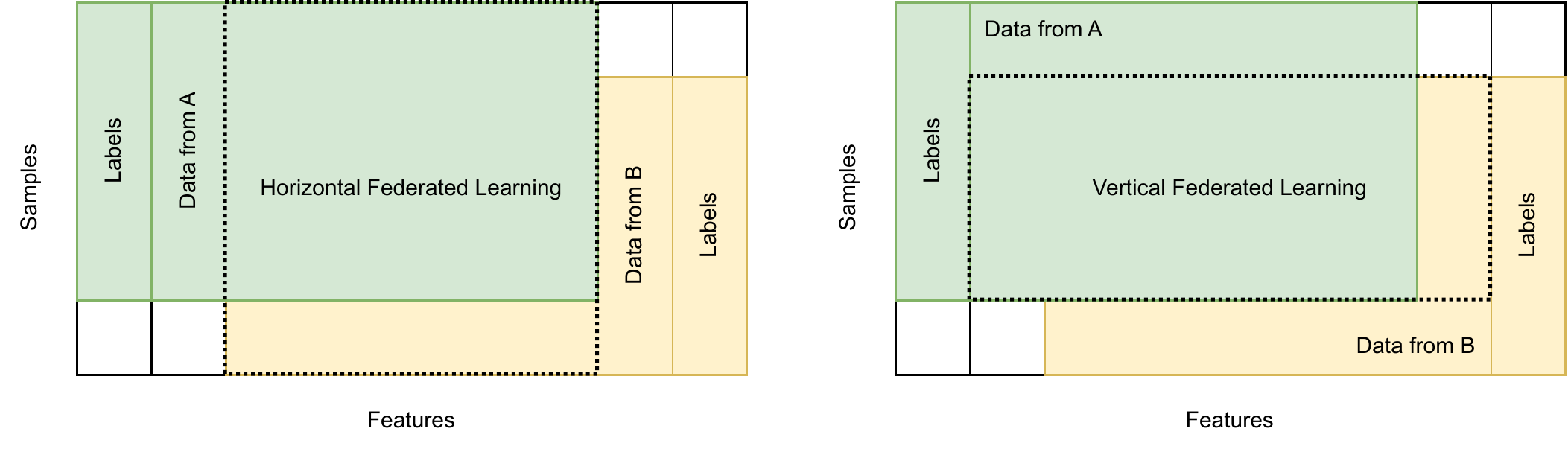}
    \caption{The comparison of horizontal and vertical federated learning.}
    \label{fig:horizontal_vertical_fl}
\end{figure*}

In several application scenarios, the datasets kept at different parties include diverse feature sets but the same entities. \add{For instance, the investment details and deposit records of a single user are usually held by a financial institution and a bank, respectively. Identifying the credit risk of an investor proves to be challenging for the financial institution due to the absence of crucial user-specific data such as deposit information. The datasets, featuring common entities but differing feature sets, fall under the category of vertically partitioned (VP)~\cite{gu2021privacy} datasets. Vertical FL is the method employed to train models across such VP datasets~\cite{trindade2021management}, as shown in Fig.~\ref{fig:horizontal_vertical_fl}. Through vertical FL, financial institutions can leverage updated gradients from banks to assess investment risks without the need for access to the user's sensitive information.}

\add{The proliferation of 5G networks and the integration of advanced hardware have led to enhanced computation and communication capabilities for IoT and edge devices~\cite{lim2020federated}. Researchers progressively deploy ML tasks on heterogeneous devices to enable intelligent human interactions while aiming to curtail communication costs~\cite{nguyen2022federated}. Nevertheless, the diversity in computational and communication capabilities among these heterogeneous devices remains inescapable. Moreover, the discrepant data sizes across nodes introduce notable discrepancies in the time required for training on each node. This often results in the generation of stale local models on slower nodes (stragglers), reducing the accuracy of the global model after aggregating.}

\begin{figure*}[ht]
    \centering
    \includegraphics[width=\linewidth]{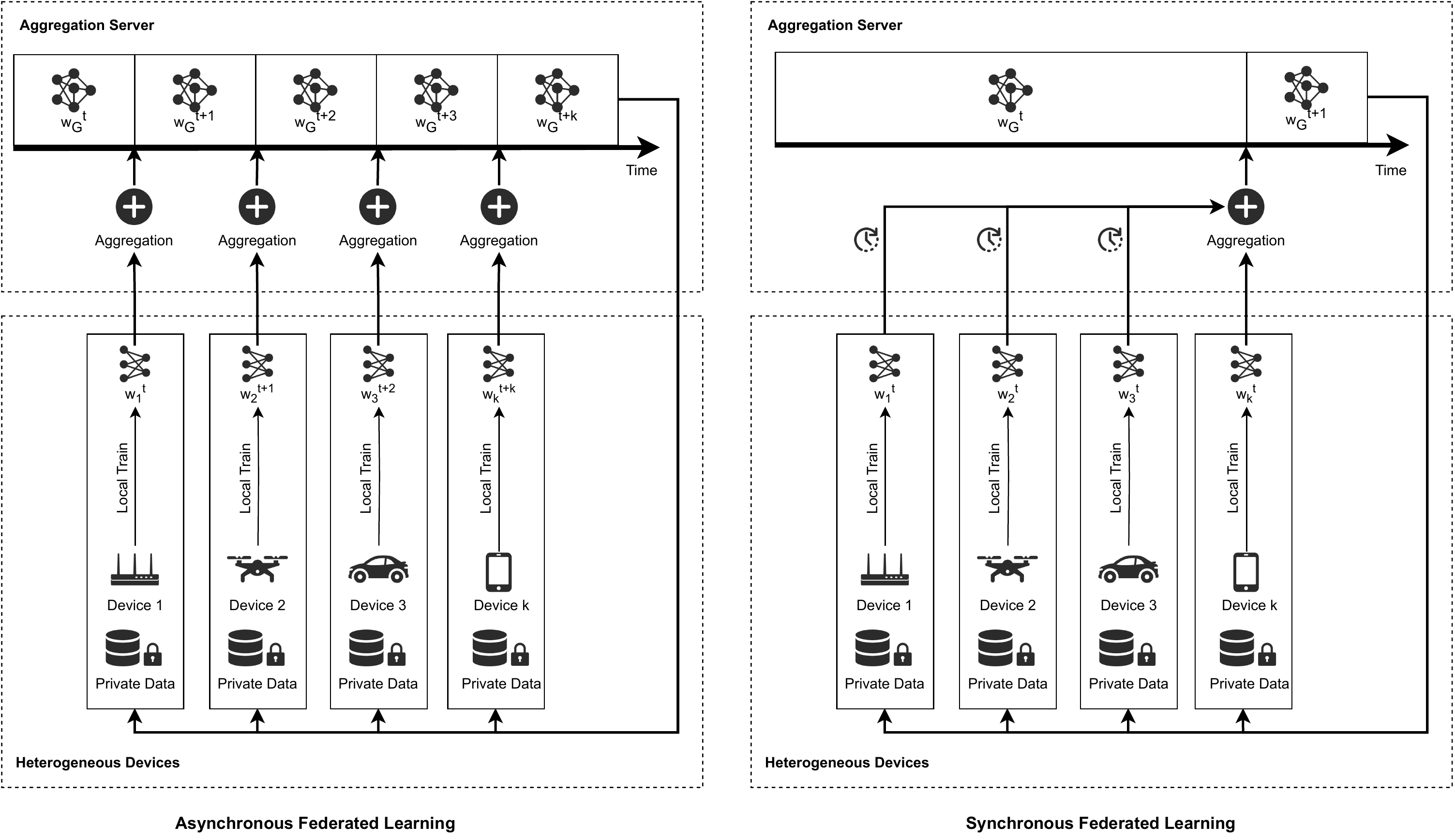}
    \caption{The comparison of workflows in asynchronous and synchronous federated learning on heterogeneous devices.}
    \label{fig:bigmap}
\end{figure*}

AFL is proposed to alleviate the impact of stale nodes and improve the efficiency of FL. \add{Within the AFL framework, global model aggregation takes place immediately once the aggregation server receives a new local model. A comparison of workflows in asynchronous and synchronous FL on heterogeneous devices is depicted in Fig.~\ref{fig:bigmap}. The principal steps of AFL are outlined as follows.}

\begin{enumerate}
\item Initialization: Similar to classic FL, the aggregation server broadcasts the initial global model $w_G^0$ to all $K$ nodes.
\item Local Model Training: \add{Nodes undertake the training of their respective local models based on the most recent global model. Due to the heterogeneity in computing capabilities among devices, the completion of local model training ($\{w_1^t, w_2^{t+1}, w_3^{t+2}, \dots, w_k^{t+k}\}$) does not occur concurrently. The local models are then sent back to the aggregation server separately.}
\item Global Model Aggregation: \add{The server aggregates the newly collected local model with the latest global model by using Eq.~\ref{eq:afl}.
\begin{equation}
w_G^{t+k} = \frac{w^{t+k-1}_G + w^{t+k}_k}{2}
\label{eq:afl}
\end{equation}
Following this process, the global model $w_G^{t+k}$ becomes available to the nodes and serves for their next iteration of local model training.}
\end{enumerate}

\add{The iteration count in AFL, denoted as $t$, increases by $1$ upon the completion of one iteration of local training by a device. The immediate model aggregation strategy in AFL reduces the waiting time for aggregation and thereby improves overall efficiency~\cite{liu2021blockchain}.}

\subsection{Blockchain}

\add{Blockchain, the backbone of Bitcoin~\cite{nakamoto2008bitcoin}, is a distributed ledger technology (DLT) that sustains the uniformity and immutability of transactional data across various nodes~\cite{xu2021lightweight}. In blockchain, nodes are responsible for maintaining the shared ledger and executing a globally unified program referred to as the smart contract. The self-verifiability and tamper-resistant attributes of the smart contract ensure the security and reliability of the shared ledger. Every node carries out the validation and execution of received transactions via the smart contract. Upon consensus attainment through the consensus algorithm, all nodes arrange transactional data into blocks and append these blocks to the shared ledger. Proof of Work (PoW)~\cite{nakamoto2008bitcoin}, Proof of Stake (PoS)~\cite{kiayias2017ouroboros}, and PBFT~\cite{sukhwani2017performance} stand as the three most prevalent consensus algorithms. Generally, a consensus algorithm with better security or fault tolerance tends to have diminished efficiency.}

Blockchain is typically treated as a distributed database for saving the models generated during the training process. Some researchers also utilize the reputation system of blockchain to motivate nodes to contribute their local models. \add{In the context of AFL, blockchain yields multiple advantages. Firstly, the immutability of shared ledgers prevents malicious node behavior, such as uploading plagiarized updated gradients~\cite{xu2023scei}. Secondly, the consensus algorithm fosters trust between unfamiliar devices, given the decentralized and unmanipulable nature of the aggregation process~\cite{qu2020blockchained}. Thirdly, the smart contract validates the authenticity of models and nodes, thus deterring malicious nodes from introducing poisoned gradients~\cite{xu2023scei}. Fourthly, the decentralized aggregation strategy on the blockchain helps the aggregation server to resist DDoS attacks and reduces the risk of single-point failures~\cite{qu2020decentralized}. However, the adoption of blockchain in AFL entails trade-offs in scalability and efficiency to a certain degree.}

\subsection{Differential Privacy}

\add{Differential privacy is a privacy-preserving technique that has experienced fast growth for over ten years. Originating from the concept of differential attack, differential privacy entails concealing a solitary sensitive data point within a particular dataset~\cite{dwork2008differential,dwork2014algorithmic}. The central objective of differential privacy is to render each data point non-discriminatory while upholding specific statistical attributes required for data analysis~\cite{cao2021data}.}

\add{Various differential privacy mechanisms have been created, each serving a crucial role in its specific application context. Prominent among these are the Laplace mechanism~\cite{dwork2014algorithmic}, exponential mechanism~\cite{dong2020optimal}, and Gaussian mechanism~\cite{zhu2020private}. By infusing controllable randomized noise, differential privacy is able to return sanitized and privacy-preserving responses to data requesters~\cite{dwork2008differential}. However, this data sanitization process in differential privacy comes at the cost of diminishing data utility~\cite{qu2018privacy}.} Thus, a parameter known as the privacy budget ($\epsilon$) is introduced to measure the balance between privacy protection and data utility.

\add{To cater to the requirements of flexible privacy protection in practical scenarios, personalized privacy protection models have been introduced. From this perspective, an index (e.g. social distance within social networks) is introduced to determine the level of personalized privacy protection~\cite{qu2021personalized}.} By fine-tuning personalized parameters, the data utility could be improved further~\cite{qu2020customizable}.

\add{While being efficient and scalable, differential privacy encounters the challenge of diminished data utility, especially when the introduced noise is subject to randomization and lacks proper control~\cite{soria2017individual}. Given that the training process in federated learning (FL) typically involves numerous local devices, sometimes numbering in the dozens or even thousands, these concerns can be mitigated by setting the mean value of the Laplace distribution to zero~\cite{wei2020federated,liu2020secure}.} Consequently, differential privacy has a great potential to be applied in FL or even AFL.

\section{Device Heterogeneity}
\label{sec:device-heterogeneity}

\add{The primary obstacle of AFL revolves around optimizing resource utilization across heterogeneous devices to enhance training efficiency. Concurrently, there exists the obstacle of stale local models resulting from device heterogeneity, a factor that is detrimental to the performance of the global model. The present study encapsulates several dimensions, encompassing node selection, weighted aggregation, gradient compression, semi-asynchronous FL, cluster FL, and model splitting. A comprehensive overview and comparison of the related work are presented in Table~\ref{table:heterogeneous-devices}.}

\begin{table*}[htpb]
    \footnotesize
    \renewcommand{\arraystretch}{1.2}
    \caption{Improve Model Performance on Heterogeneous Devices}
    \label{table:heterogeneous-devices}
    \centering
    \begin{tabular}{>{\raggedright\arraybackslash}p{0.06\linewidth}|c|>{\raggedright\arraybackslash}p{0.41\linewidth}|c|>{\raggedright\arraybackslash}p{0.2\linewidth}}
    \toprule
    & \textbf{Ref.\footnotemark[1]} & \textbf{Detail} & \textbf{Data Dist.\footnotemark[2]} & \textbf{Model \& Dataset}  \\
    \midrule
    & \cite{chen2021towards} & Heuristic greedy node selection according to local computation and communication resources. & H, I, N & CNN \& MNIST~\cite{xu2023scei}, FMNIST~\cite{xiao2017fashion}, EMNIST~\cite{cohen2017emnist} \\
    \cline{2-5}
    & \cite{zhou2021tea} & Limit the number of devices training together. & H, I, N & CNN \& FMNIST \\
    \cline{2-5}
    & \cite{hao2020time} & A prioritized node-selection function based on computing power and accuracy change. & H, I, N & CNN \& MNIST, FMNIST, CIFAR10~\cite{xu2023scei} \\
    \cline{2-5}
    & \cite{imteaj2020fedar} & Assign a trust score to each node based on its activities. & H, I, N & CNN \& MNIST \\
    \cline{2-5}
    & \cite{wu2020safa} & Select nodes with a lower crash probability. & H, I & CNN, SVM \& Boston~\cite{harrison1978hedonic}, MNIST, KDD Cup '99~\cite{stolfo2000cost} \\
    \cline{2-5}
    \multirow{-7.5}{=}{\centering \rotatebox[origin=c]{90}{Node Selection}} & \cite{hu2023scheduling} & Random, significance-based, and frequency-based scheduling are analyzed. & H, I, N & MNIST \\
    \hline
    & \cite{xie2019asynchronous} & A mixing hyperparameter that balances the convergence rate with variance reduction according to the staleness. & H, N & CNN \& CIFAR10, WikiText-2~\cite{merity2016pointer} \\
    \cline{2-5}
    & \cite{chen2019communication} & Increase the weight of recently updated local models. & H, N & CNN, LSTM \& MNIST, HAR~\cite{anguita2013public} \\
    \cline{2-5}
    & \cite{shi2020hysync} & The weight assigned to the updated gradients decreases as the staleness value increases. & H, I, N & MLP \& MNIST \\
    \cline{2-5}
    & \cite{zhou2021tea} & A caching mechanism with weighted averaging according to the staleness of the model. & H, I, N & CNN \& FMNIST \\
    \cline{2-5}
    & \cite{chen2020asynchronous} & A decay coefficient that is responsible for balancing the previous and the current model. & H, N & CNN, LSTM \& FMNIST, FitRec~\cite{ni2019modeling}, Air Qlt.~\cite{luo2019accuair}, ExtraSensory~\cite{vaizman2017recognizing} \\
    \cline{2-5}
    & \cite{xiaofeng2020asynchronous} & A duel-weighted gradient update scheme. & H, I & MLP, CNN \& MNIST, CIFAR10 \\
    \cline{2-5}
    \multirow{-12}{=}{\centering \rotatebox[origin=c]{90}{Weighted Aggregation}} & \cite{wang2021efficient} & Dynamically adjust aggregation weight of branches based on accuracy. & H, I & CNN, MBNN \& Bearing~\cite{loparo2013bearing}, Gear Fault~\cite{cao2018bearing} \\
    \hline
    & \cite{lu2020privacy} & Self-adaptive threshold computation and gradient communication compression. & H, I & MLP \& MNIST \\
    \cline{2-5}
    & \cite{li2020efficient} & A double-end sparse compression based on Top-K AllReduce sparse compression. & H, I & LR, MLP \& Insurance~\cite{dewi2019analysis}, Credit Card~\cite{yeh2009comparisons} \\
    \cline{2-5}
    \multirow{-4.5}{=}{\centering \rotatebox[origin=c]{90}{\parbox{2cm}{\centering Gradient Compression}}} & \cite{lee2021adaptive} & Three transmission scheduling algorithms for stragglers under different circumstances. & H, I & CNN \& MNIST, CIFAR10 \\
    \hline
    & \cite{hao2020time} & Local models on unselected nodes will be cached for several iterations before uploading. & H, I, N & CNN \& MNIST, FMNIST, CIFAR10 \\
    \cline{2-5}
    & \cite{wu2020safa} & Nodes are classified into three classes with tolerable nodes working asynchronously. & H, I & CNN, SVM \& Boston, MNIST, KDD Cup '99 \\
    \cline{2-5}
    & \cite{nguyen2022federated} & A private buffer on the aggregation server. & H, N & LSTM, CNN \& Sent140~\cite{go2009twitter}, CelebA~\cite{liu2015deep} \\
    \cline{2-5}
    & \cite{stripelis2021semi} & The aggregation time interval depends on the slowest node. & H, I, N & CNN \& CIFAR10, CIFAR100 \\
    \cline{2-5}
    & \cite{shi2020hysync} & The local models received in a time window are cached. & H, I, N & MLP \& MNIST \\
    \cline{2-5}
    \multirow{-8.5}{=}{\centering \rotatebox[origin=c]{90}{Semi-Asynchronous}} & \cite{so2021secure} & The server stores local models in a buffer of size K. & H, I & CNN \& MNIST, CIFAR10 \\
    \hline
    & \cite{chai2020fedat} & Nodes are clustered into faster tiers and slower tiers. & H, I, N & CNN \& CIFAR10, FMNIST, Sent140 \\
    \cline{2-5}
    & \cite{zhang2021csafl} & A metric for grouping nodes according to the gradient direction and the latency. & H, N & MCLR, LSTM \& MNIST, FMNIST, Synthetic~\cite{shamir2014communication}, Sent140 \\
    \cline{2-5}
    & \cite{xia2021vertical} & A cascade training scheme including bottom subnetworks and top subnetworks. & V & MLP, CNN \& MNIST, FMNIST, CIFAR10 \\
    \cline{2-5}
    & \cite{lee2020accurate} & Nodes are grouped based on data distributions and physical locations. & H, I, N & SR, MLP, CNN \& MNIST, FMNIST, EMNIST, CelebA \\
    \cline{2-5}
    \multirow{-9}{=}{\centering \rotatebox[origin=c]{90}{Cluster FL}} & \cite{sun2020adaptive} & Adjust the aggregation frequency among groups. & H, I & MNIST \\
    \hline
    & \cite{chen2019communication} & The parameters in shallow layers are updated more frequently than those in deep layers. & H, N & CNN, LSTM \& MNIST, HAR \\
    \cline{2-5}
    & \cite{fadlullah2020hcp} & The parameters in shallow layers are updated more frequently than those in deep layers on UAVs. & H, I & CNN \& Real-World Movie Ratings \\
    \cline{2-5}
    \multirow{-4}{=}{\centering \rotatebox[origin=c]{90}{\parbox{1cm}{\centering Model Splitting}}} & \cite{wang2021efficient} & Allow nodes to select a branch of the global model based on local data distribution. & H, I & CNN, MBNN \& Bearing, Gear Fault \\
    \bottomrule
    \multicolumn{5}{l}{\textsuperscript{1} \footnotesize{Reference paper that belongs to the specific group.}} \\
    \multicolumn{5}{l}{\textsuperscript{2} \footnotesize{Data distribution across nodes. H: horizontal, V: Vertical, I: IID, N: non-IID.}} \\
    \end{tabular}
\end{table*}

\subsection{Node Selection}

\add{Numerous algorithms for node selection are proposed to improve the training efficiency of AFL across heterogeneous devices. In contrast to classic FL selecting nodes with more training data, AFL leans towards prioritizing nodes with heightened resilience and computational capacity. However, it is challenging to strike a balance between robustness and overfitting of the global model.}

For instance, in~\cite{chen2021towards}, the authors present a heuristic greedy node selection strategy that iteratively selects heterogeneous IoT nodes to participate in global learning aggregation based on their local computing and communication resources. Experiments are conducted on both IID and non-IID datasets to verify the effectiveness of their approach. 
Apart from that, considering the large number of edge devices involved, in~\cite{zhou2021tea}, the authors limit the number of devices training simultaneously in the AFL network. A limit-size cache with a weighted averaging mechanism is introduced onto the server to reduce the impact of model staleness. Experiment results back up the improved convergence speed and model accuracy. \add{These schemes are simple to improve aggregation efficiency but lack the sense of non-IID data across different nodes.}

In order to select nodes more reasonably, in~\cite{hao2020time}, a prioritized node-selecting function is designed according to the computing power and accuracy change of local models on each node. Other unselected nodes continue the iterations locally at the same time. As a result of the node-selecting function, the experiment results show a higher accuracy growth rate with a faster convergence speed. \add{Nevertheless, this prioritized node-selection method does not consider the device unreliability of IoT devices. Thus, the authors of~\cite{imteaj2020fedar} propose an idea that assigns a trust score to each node based on its activities.} ML tasks with resource requirements and a minimum trust score are published in the FL network. Any candidates who do not meet the task requirement are filtered out before the training round begins. Clients who complete tasks will be rewarded, while those who do not will have their trust value decreased.
Similarly, in~\cite{wu2020safa}, a node with a lower probability to crash is more likely to be selected in an iteration. The straggling nodes that training models that are too stale will be tagged as deprecated and forced to synchronize with the server. The tolerable nodes are those training on the acceptable stale models, who work asynchronously with the server. After updated gradients from a fraction of nodes are received, the central server ends a round of training. As a result, the waste of computation resources is minimized, and communication expenses are kept at a relatively low level.

\add{For a more comprehensive assessment of diverse device scheduling and update aggregation strategies, a study conducted by the authors in~\cite{hu2023scheduling} involves experiments encompassing both IID and non-IID datasets. These experiments are conducted across a spectrum of computational resources and training data distributions, considering scenarios where a subset of IoT devices is permitted to upload local models.} Specifically, the device scheduling policies include random scheduling, significance-based scheduling, and frequency-based scheduling; the update aggregation policies include equal weight aggregation and age-aware aggregation. The simulation results demonstrate that the random scheduling policy outperforms others while training on non-IID datasets. Besides, an appropriate age-aware aggregation policy performs better.

\subsection{Weighted Aggregation}

\add{Numerous weighted aggregation algorithms have been introduced to lessen the influence of slow devices and increase learning efficiency. In conventional FL, weighted aggregation aims to amplify the influence of local models trained with more data. However, in AFL, the objective shifts toward alleviating the effects of stale local models, which does not exist in classic FL.}

\add{One primary idea is introducing a parameter that accounts for staleness, which reduces the influence of stale local models and elevates the influence of the most recent local models during the aggregation procedure. There are several papers that adopt this method. For example, in~\cite{xie2019asynchronous}, a mixing hyperparameter is introduced based on staleness to balance the convergence rate and variance reduction.} The experiments conducted on CIFAR-10 and WikiText-2 validate both fast convergence and staleness tolerance.
In~\cite{chen2019communication}, a temporally weighted aggregation strategy is proposed, which increases the weight of recently updated local models when aggregating on shallow and deep layers. Experiment results on CNN and LSTM neural networks show that the global model accuracy and convergence are improved. 
Another time-based weighted aggregation algorithm is proposed in~\cite{shi2020hysync}. The weight assigned to the updated gradients decreases as the staleness value increases. 
Similarly, in~\cite{zhou2021tea}, a staleness-based weighted aggregation algorithm with cache is proposed.
In~\cite{chen2020asynchronous}, a decay coefficient is proposed with similar effects, balancing the previous and current models. With the dynamic learning step size, the nodes with more data or poor communication status are compensated. Experiments across three real-world datasets are conducted with results showing that their scheme converges fast and enables higher model accuracy.

\add{Nevertheless, the aforementioned methods only focus on the staleness of local models, which is a one-sided view. Thus, a duel-weighted gradient updating strategy is proposed in~\cite{xiaofeng2020asynchronous}, which takes into account the size of the dataset as well as the similarity between the local and global gradients.} The updated gradients submitted by edge devices are aggregated after the duel-weight correction. The experiment results reveal that the model accuracy remains high even after gradient compression.

\add{Apart from setting more factors of weighted aggregation, to enable more fine-grained weighted aggregation, an idea is to aggregate branches in a model with weights.}
In~\cite{wang2021efficient}, the global model is split into branches with the aggregation procedure transformed into a branch-weighted merging process. The aggregation weight is dynamically adjusted depending on the training accuracy of all nodes to prevent the global model from overfitting to nodes that upload gradients frequently. To evaluate the effectiveness of the proposed scheme, a prototype is implemented on heterogeneous devices based on two industrial cases: (1) Fault diagnosis of motor bearings and (2) Fault diagnosis of the gearbox. The experiment results demonstrate that their scheme converges faster, achieves higher accuracy, and consumes less energy than the classic CNN model.

\subsection{Gradient Compression}

\add{As gradient compression stands as a widely applicable tactic for improving the efficiency of FL, its incorporation into AFL commonly aims to achieve an additional reduction in communication expenses. Nevertheless, AFL introduces new challenges to gradient compression, primarily within the resource-restricted computing environments of edge and IoT devices, along with a higher frequency of aggregation operations. Specifically, the disparity in computational capabilities among nodes is much more significant. Besides, AFL incurs larger computational demand on the server side compared to classic FL due to the intensified frequency of aggregation and compression operations. To address these challenges, several efficient gradient compression algorithms tailored for AFL have been presented.}

\add{For instance, in~\cite{lu2020privacy}, two sub-modules are presented for self-adaptive threshold gradient compression: (1) self-adaptive threshold computation and (2) gradient communication compression.} The former is in charge of computing the threshold based on recent parameter changes, while the latter is in charge of compressing redundant gradient communications based on the threshold. The accuracies of the generated models after gradient compression are verified when training the MLP model on the MNIST dataset. Besides, the proposed scheme allows the node to join or quit freely, which is suitable to highly mobile edge computing scenarios. Another similar paper is published in~\cite{xiaofeng2020asynchronous} by the same authors.

\add{From the perspective of vertical FL, in~\cite{li2020efficient}, based on the Top-K AllReduce sparse compression technique, the authors present a double-end sparse compression algorithm~\cite{aji2017sparse}.} Specifically, the compression process happens on both the server and local sides to reduce the transmission cost. Experiment results demonstrate that $86.90\%$ of information exchange is minimized during the training process, revealing that their scheme is suitable for edge computing scenarios with low-bandwidth or metered networks. Furthermore, the training data is protected securely against gradients leakage attacks~\cite{zhu2020deep}.

Another approach to improve communication efficiency is to design a new communication protocol that more efficiently schedules model upload and download. For example, in~\cite{lee2021adaptive}, three transmission scheduling algorithms that account for slow nodes are proposed to improve the efficiency of AFL in wireless networks, where statistical information regarding uncertainty is known, unknown, or limited. The experiment results show their outperformance in terms of accuracy, convergence speed, and robustness.

\subsection{Semi-Asynchronous FL}

\add{In AFL, the inclusion of stale local models from slow nodes during aggregation diminishes the accuracy of the global model to a certain extent. To alleviate the effects of these slow devices, semi-asynchronous FL schemes have been introduced. Typically, semi-asynchronous FL serves as a hybrid approach that combines elements from both classic FL and AFL, where the aggregation server captures and stores local models that arrive earlier, subsequently aggregating them following a certain timeframe. Depending on the magnitude of staleness, the subsequent arrivals of local models either take part in the following training rounds or are discarded. Notably, the aggregation frequency of semi-asynchronous FL falls between that of AFL and classic FL. Similar to classic FL, a training round is defined as the process spanning from one global aggregation to the next.}

\add{For example, in~\cite{hao2020time}, a priority function is introduced to accurately select nodes with large amounts of data or high computation power.} Meanwhile, local models on unselected nodes will be cached for a specific number of iterations before being submitted to the aggregation server. Besides, a restriction on the local training round number is set to prevent specific nodes from being unselected for a lengthy period of time, leading the global model to overfit certain nodes. The effectiveness of the scheme is evidenced by experiments conducted on IID and non-IID datasets. \add{However, the restriction in this scheme will amplify the influence of stale local models after numerous aggregations.}

On the contrary, a cache-based lag-tolerant mechanism on the aggregation server is introduced in~\cite{wu2020safa} to mitigate the impacts of stragglers, crashes, and model staleness. In their scheme, all nodes are classified into three categories: up-to-date, deprecated, and tolerable. Only the up-to-date and deprecated nodes are forced to synchronize with the server, while the tolerable nodes work asynchronously. The nodes will be labeled picked, undrafted, or crashed after training. Specifically, local models from undrafted nodes are not aggregated in this round but retained in the cache for aggregation with local models in the next round. As a result, the tradeoff between faster convergence and lower communication overhead is properly addressed, which is verified by experiments.
Similarly, a private buffer on the aggregation server holding a certain number of model updates is designed in~\cite{nguyen2022federated}, with convergence ensured by math. To evaluate the scalability and efficiency of their scheme under various staleness distributions, the authors train an LSTM classifier on text and image classification tasks. The results reveal that their approach is more resistant to diverse distributions and converges faster than classic synchronous and asynchronous FL schemes. \add{However, the aforementioned schemes do not take the security of the cache into account. To further improve cache security, a scheme that adopts a secure buffer on the server is proposed in~\cite{so2021secure}, where a secure aggregation protocol is designed to prevent the server from learning any information about the local updates.}

\add{From the perspective of time, the authors in~\cite{stripelis2021semi} aggregate local models at a specific time interval determined by the slowest node.} More exact control over the training nodes is allowed, especially in edge computing networks with non-IID data distribution. The authors then compare classic synchronous, asynchronous, and semi-synchronous schemes across heterogeneous devices in experiments. The results show that their approach is faster and more accurate than other schemes.

\add{For AFL, there is a good chance that a significant number of local model updates come in a short period of time. Considering this issue, in~\cite{shi2020hysync}, after caching the first several local models received within a given time window, a synchronous aggregation strategy is adopted.} The experiment results reveal that compared with the classic FL scheme, the time window enables their scheme many more nodes.

\subsection{Cluster FL}

\add{Clustered FL is an approach geared towards augmenting training efficiency through the formation of clusters comprising devices exhibiting similar performance, functionalities, or datasets. The asynchronous update strategy has the potential to bring advantageous to inner-group updates, inter-group updates, or both}

For instance, an idea is grouping nodes into tiers based on their response latency~\cite{chai2020fedat}. Faster tiers are responsible for faster convergence, while slower tiers aid in the model accuracy improvement. Furthermore, a polyline-encoding-based compression algorithm is adopted in their scheme to improve communication efficiency. Experiments are conducted across multiple datasets and models, confirming that their scheme has a low communication cost and high prediction accuracy. \add{However, grouping nodes by only considering the factor of response latency is circumscribed.
By contrast, in~\cite{zhang2021csafl}, a grouping metric is proposed, where the gradient direction and the latency of model update are taken into account.} The local update latency is composed of computation latency and communication latency. Experiments conducted on four imbalanced non-IID datasets assess the improvement in test accuracy.

\add{From the aspect of grouping architecture, in~\cite{xia2021vertical}, a cascade training scheme with bottom and top subnetworks is proposed to fully exploit all horizontally partitioned labels.} Specifically, the bottom subnetworks are responsible for extracting embedding vectors from features, while the top subnetworks are for prediction. The nodes in FL are classified into three types, including active party, passive party, and collaborator. Each active party is connected to other passive parties so that it is able to gather embedding vectors and return gradients to them. The collaborator is connected to all active parties in order to aggregate the returned gradients. The experiment results reveal that their scheme effectively addresses the straggler problem with minimum performance loss.

\add{Another advantage of cluster FL is improving communication efficiency. For example, in~\cite{lee2020accurate}, nodes are grouped based on data distributions and physical locations to reduce global model loss and communication delay.} The authors designed a control algorithm that reduces communication costs while examining the convergence of the proposed scheme in IID settings. The outperformed accuracy and efficiency of their scheme are evidenced by the experiment results. \add{Additionally, Cluster FL allows different groups of nodes to aggregate at different frequencies, which also reduces communication costs. For example, in~\cite{sun2020adaptive}, the authors adaptively modify the aggregation frequency among groups to minimize the loss of FL.} Under an environment with limited resources, a dynamic trade-off between computation and communication cost is formulated by Markov Decision Process (MDP) and optimized by deep reinforcement learning (DRL). Numerical results validate the accuracy, convergence, and energy-saving features of their proposed scheme.

\subsection{Model Splitting}

\add{Following the splitting of the deep neural network model, each node is responsible for training a certain part rather than the whole model. Thus, the model-splitting strategy curtails the number of parameters necessitating transmission, consequently leading to an enhancement in communication efficiency. Upon integrating the model-splitting strategy into AFL, nodes bypass the need to await other nodes and fully utilize their computing resources to train the model for subsequent rounds. Therefore, the model-splitting strategy expedites the convergence of the global model to a certain extent.}

\add{For instance, in~\cite{chen2019communication}, a layerwise asynchronous model updating strategy is proposed, in which shallow layer parameters are updated more frequently than deep layer parameters.} When aggregating, the most recently updated local models have the highest weight with the help of timestamps. The experiment results support the improved communication cost and model accuracy of the proposed scheme.
In~\cite{fadlullah2020hcp}, a similar idea is achieved by using cache and communication capabilities on UAVs and terrestrial base stations. The parameters in shallow layers are updated more frequently than those in deep layers. To predict the content caching placement, the proposed scheme employs a two-stage AFL algorithm. The efficiency of the proposed scheme is validated by experiments conducted on real-world datasets and numerical analysis.

\add{Apart from splitting models into deep and shallow layers, an approach is to divide the global model into branches according to the sample category~\cite{wang2021efficient}.} The splitting process involves acquiring a branch from the entire model. The aggregation process is performed on branches with different weights dynamically adjusted by the aggregation server. Besides, it allows nodes to select parts of the model according to local data distribution and update asynchronously reduces calculation and communication costs, enhancing FL efficiency.

\add{In contrast to the node selection strategy, the model-splitting strategy reduces the computational demands placed upon nodes, affords greater flexibility in updating distinct layers of the global model, and alleviates biases present within the global model. Nonetheless, the extendability of this strategy to different models is constrained, primarily due to the requisite implementation of customized splitting and aggregation algorithms for every model across various datasets.}

\section{Data Heterogeneity}
\label{sec:data-heterogeneity}

In practice, the data across nodes is usually non-IID. Besides, the amount of data distributed on each node is always imbalanced. \add{Consequently, the frequent uploading of models on particular nodes has the potential to attract divergence to the global model and result in overfitting to specific datasets.}

\subsection{Non-Independent and Identically Distributed Data}

\add{The presence of non-IID data among nodes tends to cause a biased global model in AFL. To tackle the issues posed by this non-IID data, the research domain primarily encompasses four avenues of exploration, including constraint terms for aggregation, clustered FL, distributed validation strategy, and mathematically optimizing parameters.}

\add{A typical example of constraint terms for aggregation is~\cite{chai2020fedat}, where a constraint term is presented to limit local updates to be closer to the global model.} Besides, nodes with similar updating frequencies are grouped into the same tier through synchronous and asynchronous training strategies to prevent local models from diverging. The effectiveness of the scheme is supported by mathematical analysis and experiment results.

\add{Clustered FL is also a strategy to alleviate the effects of divergent data distributions by grouping training nodes. In~\cite{sattler2020clustered}, the geometric properties of the FL loss surface are used to group nodes into clusters. The quality of clusters is ensured by math and validated by experiments. In~\cite{lee2020accurate}, the data distribution on nodes in a group is optimized to be closer to the global data distribution. To better group nodes in general situations, the authors in~\cite{zhang2021csafl} propose a spectral clustering approach, where nodes are grouped based on an affinity matrix derived from model update latency and direction. Non-IID dataset settings are also applied in their experiments, with the results showing that their scheme enhances test accuracy and convergence speed.}

\add{An example of distributed validation strategy is~\cite{stripelis2020accelerating}, where the authors propose a distributed validation scheme that evaluates model performance across nodes.} A small percentage ($5\%$) of local training data samples is reserved on each node to evaluate models from other nodes. As a result, a better generalized global model is obtained. By adopting both synchronous and asynchronous communication protocols, models trained on heterogeneous data and compute environments demonstrate the superior performance of the proposed scheme.

\add{From the perspective of mathematical analysis, in~\cite{tian2021towards}, a training strategy with pre-determined initial weight parameters is proposed to mitigate the global model divergence.} By using the Taylor Expansion formula, higher precision gradients are achieved on their AFL scheme, which is validated by experiments conducted across many real-world datasets. Another mathematical solution of non-IID data is choosing optimal hyper-parameters for the novel two-stage training strategy in AFL~\cite{chen2021fedsa}.

\add{In addition to the aforementioned methods, a number of research papers analyze the effects of stale or imbalanced non-IID data on AFL. For example, in~\cite{diwangkara2020study}, an AFL scheme is proposed, where the authors focus on how staleness and data imbalance affect AFL by performing various levels of experiments.} The results reveal that AFL works effectively on balanced data distribution when the server update frequency is unequal. Considering the effects of smooth strongly convex and smooth nonconvex functions when the data distribution is non-IID, the authors in~\cite{avdiukhin2021federated} investigate the convergence theoretically and conduct several experiments. The results show that AFL has the same convergence rate as traditional FL while lowering communication requirements. By implementing the AFL scheme and conducting experiments on six Raspberry Pi 3B+ devices, the authors in~\cite{yang2020prototyping} investigate the impact of heterogeneous devices. The results of experiments conducted on the MNIST dataset with non-IID data distribution reveal that AFL outperforms classic FL, especially when computing resources and input data sizes are disparate.

\subsection{Vertical Distributed Data}

\add{In contrast to horizontal FL, vertical FL focuses on datasets where distinct subsets of features are spread across various nodes, as elaborated upon in Section~\ref{sec:bg_fl}. Given that the creation of the global model depends on the aggregation of local models, there exists a requirement for collaborative updating of these local models. Such skewed feature distribution and heightened model interdependence consequently present challenges to vertical AFL. To tackle these challenges, the state-of-the-art schemes mainly focus on improving communication and training efficiency.}

\add{One research direction is to improve the communication efficiency of vertical FL.} In~\cite{chen2020vafl}, apart from the flexible FL algorithm that allows random client participation, the authors utilize a local embedding model for each client to convert raw input to compact features, reducing the communication parameters in AFL. The feasibility and effectiveness of the proposed scheme are confirmed by rigorous convergence analysis and numerical experiments on multiple datasets.

\add{In addition to communication efficiency, the training efficiency also needs to be optimized in vertical FL.} For example, the authors in~\cite{gu2021privacy} propose an asynchronous federated stochastic gradient descent (AFSGD-VP) algorithm with two variance reduction variants: stochastic variance reduced gradient (SVRG) and SAGA~\cite{defazio2014saga}. When the objective function is strongly convex, the convergence rate of AFSGD-VP is derived. Besides, the security and complexity of the proposed algorithm are provided. Experiment results on several vertical distributed datasets verify the theoretical analysis and prove the efficiency of their proposed scheme. \add{Apart from the stochastic gradient descent, backward updating is also a key stage of training.} A vertical AFL scheme with a backward updating mechanism and a bilevel asynchronous parallel architecture is proposed in~\cite{zhang2021secure}. Specifically, the backward updating mechanism enables all parties to update the model in a secure manner. The bilevel asynchronous parallel architecture improves the efficiency of the backward updating process. As the name implies, the parallel architecture is divided into two levels: the inner-party parallel between active parties and the intra-party parallel within each party. Both levels of the update are performed asynchronously to improve efficiency and scalability. The authors demonstrate the feasibility and security of the proposed strategy through theoretical and security analysis. Experiments on real financial datasets are conducted, whose results demonstrate efficiency, scalability, and losslessness.

\add{Upon the aforementioned strategies, a hybrid approach that improves both training and communication efficiency is proposed.} In~\cite{li2020efficient}, the authors propose a vertical AFL scheme with gradient prediction and double-end sparse compression algorithm. In particular, the gradient prediction presents the timely renewal of participants by using second-order Taylor expansion, reducing training time while retaining a sufficient degree of accuracy. The double-end sparse compression algorithm reduces the amount of data exchanged across the network during the training process. Experiment results obtained by training models on two public datasets reveal the outperformed efficiency of the scheme without degrading the accuracy and convergence speed.

\section{Privacy and Security on Heterogeneous Devices}
\label{sec:privacy-security}

\add{While FL is initially introduced to protect the privacy of local training data, new attack vectors have emerged, causing privacy concerns, such as membership inference attack~\cite{shokri2017membership}, property inference attack~\cite{melis2019exploiting}, model inversion attack~\cite{fredrikson2015model}, and deep leakage from gradients attack~\cite{zhu2020deep}. Several attacks, like poisoning attacks or backdoor attacks, pose a significant threat to the integrity of the global model. Remedial measures for privacy and security concerns within FL include leveraging techniques like differential privacy and blockchain. Since AFL operates as a variant of FL, it is vulnerable to these attacks when training models across heterogeneous devices. However, the current solutions addressing privacy and security issues are often computationally demanding, rendering them less viable for deployment on resource-constrained heterogeneous devices and time-sensitive application scenarios. To confront these obstacles, many studies have come forward, presenting flexible differential-privacy models or highly efficient blockchain-based solutions tailored to AFL. A comparative analysis of these papers is presented in Table~\ref{table:attack-resistance}.}

\begin{table*}[ht]
    \footnotesize
    \renewcommand{\arraystretch}{1.2}
    \caption{Attack Resistance Comparison on Heterogeneous Devices}
    \label{table:attack-resistance}
    \centering
    \begin{tabular}{c|c|c|c|c|c|c}
    \toprule
    \textbf{Ref.\footnotemark[1]} & \textbf{BKG\footnotemark[2] Attack} & \textbf{Collusion Attack} & \textbf{Inference Attack} & \textbf{Poisoning Attack} & \textbf{Byzantine Attack} & \textbf{DDoS Attack} \\
    \midrule
    \cite{chen2020vafl} 
    & $\surd$ & $\surd$ & $\surd$ & $\times$ & $\times$ & $\bigcirc$ \\
    \cite{lu2019differentially} 
    & $\surd$ & $\surd$ & $\surd$ & $\bigcirc$ & $\bigcirc$ & $\bigcirc$ \\
    \cite{li2019asynchronous} 
    & $\surd$ & $\surd$ & $\surd$ & $\times$ & $\times$ & $\bigcirc$ \\
    \cite{van2020asynchronous} 
    & $\surd$ & $\surd$ & $\surd$ & $\times$ & $\times$ & $\bigcirc$ \\
    \cite{lu2020communication1} 
    & $\times$ & $\times$ & $\times$ & $\bigcirc$ & $\bigcirc$ & $\bigcirc$ \\
    \cite{lu2020communication2} 
    & $\times$ & $\times$ & $\times$ & $\times$ & $\times$ & $\surd$ \\
    \cite{lu2020blockchain} 
    & $\times$ & $\times$ & $\times$ & $\bigcirc$ & $\bigcirc$ & $\surd$ \\
    \cite{liu2021blockchain} 
    & $\times$ & $\times$ & $\times$ & $\surd$ & $\bigcirc$ & $\bigcirc$ \\
    \cite{feng2021blockchain} 
    & $\times$ & $\times$ & $\times$ & $\surd$ & $\bigcirc$ & $\bigcirc$ \\
    \cite{yuan2021chainsfl} 
    & $\times$ & $\times$ & $\times$ & $\bigcirc$ & $\bigcirc$ & $\bigcirc$ \\
    \cite{xu2021bafl} 
    & $\times$ & $\times$ & $\times$ & $\surd$ & $\bigcirc$ & $\surd$ \\
    \cite{xu2022efficient} 
    & $\times$ & $\times$ & $\times$ & $\surd$ & $\bigcirc$ & $\surd$ \\
    \cite{xu2022bass} 
    & $\bigcirc$ & $\bigcirc$ & $\bigcirc$ & $\surd$ & $\bigcirc$ & $\surd$ \\
    \bottomrule
    \multicolumn{7}{l}{$\surd$ Fully resistant to the attack; $\bigcirc$ Partially resistant to the attack; $\times$ Not resistant to the attack.} \\
    \multicolumn{7}{l}{\textsuperscript{1} \footnotesize{Reference paper that belongs to the specific group.}} \\
    \multicolumn{7}{l}{\textsuperscript{2} \footnotesize{Background Knowledge Attack.}} \\
    \end{tabular}
\end{table*}

\subsection{Privacy on Heterogeneous Devices}

\add{Differential privacy stands as a promising methodology embraced within various AFL schemes to protect the privacy of local models, consequently mitigating the risk of local training data leakage on heterogeneous devices.}

\add{For instance, in~\cite{chen2020vafl}, a flexible FL scheme with differential privacy is proposed to avoid disclosing local training data.} Each node employs Gaussian differential privacy to achieve a better trade-off between data privacy and data utility. \add{However, a universal differential privacy setting lacks flexibility across heterogeneous devices. Consequently, in~\cite{lu2019differentially}, the authors proposed an AFL scheme adopting local differential privacy for secure resource sharing in vehicular networks.} Particularly, a tree-based gradient descent model is adopted on nodes to achieve high global model accuracy in a short amount of time. To protect the privacy of local models, a distributed local model updating approach with Gaussian noise is introduced to nodes in the regression tree. By offering rewards, nodes are encouraged to provide good models, thereby accelerating the convergence process. The experiment carried out on three real-world datasets demonstrates the high accuracy and efficiency of the scheme.

\add{Apart from investigating personalized differential privacy, some researchers focus on the balance between model utility and privacy protection. For example, in~\cite{li2019asynchronous}, the convergence of AFL while adopting differential privacy is analyzed.} Based on the analysis, a multi-stage adjustable algorithm is proposed to optimize the trade-off between model utility and privacy by dynamically changing the noise size and the learning rate. Experiments are conducted on edge servers and a cloud server with three different ML models, including logistic regression (LR), support vector machine (SVM), and convolutional neural network (CNN). The results reveal that MAPA achieves high model utilities and accuracy at the same time.
Furthermore, in~\cite{van2020asynchronous}, differential privacy is introduced into AFL by adding Gaussian noise. The authors begin AFL training with a high learning rate and gradually reduce it to achieve optimum accuracy. The theoretical analysis and simulation results prove that their scheme reduces the network communication cost on heterogeneous devices.

\subsection{Security on Heterogeneous Devices}

\add{When training ML models on heterogeneous devices, the blockchain is typically leveraged as a secure distributed database to ensure the secure storage or transmission of local models within AFL. The advantages of the integration include privacy protection and trust promotion among heterogeneous devices. For example, in~\cite{lu2020communication1}, blockchain and digital twin edge network are integrated to store all local gradient updates in AFL.} Specifically, the blockchain is adopted to track the aggregation progress by maintaining a global iteration index in AFL. A lightweight DPoS-based verification mechanism is developed, where stakes are earned based on the computing contribution to the global model. The mechanism is accomplished through the verification algorithm, which verifies the quality of the models against the historical model. In addition, a reinforcement learning-based algorithm is designed for efficient user scheduling and bandwidth allocation. A series of experiments are conducted to evaluate the performance of the scheme in terms of learning accuracy and resource cost. Another similar idea proposed by the same authors in~\cite{lu2020communication2} is to integrate blockchain, FL, and an asynchronous model update scheme in digital twin edge networks. The objective of lowering communication costs includes two parts: reducing transmission data size and optimizing communication resource allocation. Finally, the communication resource allocation approach is implemented by using deep neural networks. Numerical results reveal that this scheme improves communication efficiency and reduces the cost of resources. \add{Apart from improving communication efficiency, some researchers focus on improving learning efficiency when integrating AFL with blockchain.} The authors of~\cite{lu2020blockchain} present an AFL scheme coupled with the Directed Acyclic Graph (DAG) blockchain for the Internet of Vehicles (IoV). The participating nodes are selected by Deep Reinforcement Learning (DRL) to improve the learning efficiency. Besides, a two-stage verification mechanism is developed, which comprises periodic validation of blockchain transactions and validation of local model quality. Experimental results show the excellent learning accuracy and rapid convergence speed of the scheme.

\add{Another advantage of the integration is mitigating the risk of single-point failure caused by the centralized aggregation server. For instance, a blockchain-based AFL scheme with a staleness coefficient is proposed in~\cite{liu2021blockchain}.} Specifically, the staleness coefficient reduces the contribution from the latency device to the global model by comparing the version of the global model with the stale local model. The Proof-of-Work (PoW) consensus algorithm is adopted, where the miners are responsible for generating candidate blocks that include trained models. The block generation rate is positively correlated with the forking frequency. The experiments carried out on a variety of IoT devices demonstrate high accuracy on both horizontal and vertical FL frameworks. Another similar idea with a staleness coefficient is proposed in~\cite{xu2021bafl, xu2022efficient}. Instead of using PoW, a committee-based consensus algorithm is adopted to improve efficiency further. The convergence speed and model accuracy are both validated by experiments on heterogeneous devices. \add{Instead of staleness coefficient, some researchers pay attention to the architecture of the integration. For instance, to mitigate the risk of single-point failure and malicious nodes attacks, the authors in~\cite{yuan2021chainsfl} propose a two-layer blockchain-driven FL framework composed of multiple Raft-based shard networks (layer-1) and a DAG-based main chain (layer-2).} Layer-1 is a small group for information exchange, while layer-2 is responsible for storing and sharing models trained by layer-1 asynchronously. Furthermore, to avoid the impact of stale models, a virtual pruning procedure with a specific waiting time is presented. Models not approved by other models for a long time or with low accuracy will be pruned from the DAG blockchain. The experiment results show that this scheme is resilient against malicious nodes while maintaining acceptable convergence rates.

\add{Additionally, the reputation of nodes is an important factor to be considered to improve the stability and security of AFL, which is easily handled by the blockchain by its built-in reputation and reward systems. For example, in~\cite{feng2021blockchain}, a blockchain-based AFL scheme is proposed, where an entropy weight method determines the participant rank by the proportion of local models trained on nodes.} The metrics are all maintained in the blockchain, including the training time, training sample size, local update correlation, and global update cheating times.
The resource cost and training efficiency are well balanced by optimizing local training delays and the block generation rate. The experiment results show the superiority of the scheme in terms of efficiency and preventing poisoning attacks.

\add{Moreover, some researchers investigate the effect of the gradient compression algorithm on the integration of blockchain and AFL, aiming at improving security and efficiency simultaneously. For example, in~\cite{xu2022bass}, the authors propose a SignSGD-based asynchronous federated learning paradigm (BASS) that only uploads the signs of the gradients of local models to the aggregation server, mitigating the risks of poisoning attacks. Theoretically, their paradigm is able to resist both privacy and security attacks at the cost of model convergence speed or accuracy. However, they fails to validate the performance of their paradigm in privacy preserving.}

\section{Applications on Heterogeneous Devices}
\label{sec:applications}

\add{AFL is adopted in various application scenarios, offering an efficient and adaptable training procedure on heterogeneous devices while upholding the privacy of local training data. The application scenarios of AFL and correlated research endeavors are summarized and compared in Table~\ref{table:application-scenarios}.}

\begin{table*}[ht]
    \footnotesize
    \renewcommand{\arraystretch}{1.2}
    \caption{The Applications on Heterogeneous Devices}
    \label{table:application-scenarios}
    \centering
    \begin{tabular}{>{\centering\arraybackslash}p{0.06\linewidth}|c|c|c|>{\arraybackslash}p{0.45\linewidth}}
    \toprule
    & \textbf{Ref.\footnotemark[1]} & \textbf{Use Case} & \textbf{FL Client \& Server} & \textbf{Key Contributions} \\
    \midrule
    & \cite{lu2020blockchain} & IoV & Vehicle \& MBS & Improve reliability and efficiency of data sharing and traffic prediction among vehicles. \\
    \cline{2-5}
    & \cite{lu2019differentially} & IoV & Vehicle \& MBS, RSU & Efficiently and securely allocate resources for vehicles. \\
    \cline{2-5}
    & \cite{lu2020communication1} & IoV & IoT \& Base Station & Provide high-quality services with optimized network and resources allocation. \\
    \cline{2-5}
    & \cite{zhang2021real} & IoV & Vehicle \& Server & Predict real-time steering wheel angle for autonomous driving. \\
    \cline{2-5}
    & \cite{lu2020privacy} & IoV & Camera \& Server & Monitor, predict, and adjust traffic by controlling signal lights. \\
    \cline{2-5}
    & \cite{xiaofeng2020asynchronous} & IoV & Camera \& Server & Monitor, predict, and adjust traffic by controlling signal lights. \\
    \cline{2-5}
    & \cite{yang2021privacy} & UAV & Mobile \& UAV & Provide efficient communication and computation services for ground mobile devices in outdoor events. \\
    \cline{2-5}
    & \cite{fadlullah2020hcp} & UAV & Mobile \& UAV & Provide a content caching system in UAV networks to extend the service coverage and reduce the communication delay of the 6G network. \\
    \cline{2-5}
    \multirow{-11}{=}{\centering \rotatebox[origin=c]{90}{Smart Transportation}} & \cite{imteaj2020fedar} & Mobile Robot & Robot \& Server & Improve the performance of the ML models on mobile robots with low communication costs. \\
    \hline
    & \cite{ma2021asynchronous} & Fault Diagnosis & Edge \& Server & Track actual system changes in real-time and improve the diagnostic rate of the devices. \\
    \cline{2-5}
    & \cite{tian2021towards} & Fault Diagnosis & Edge \& Server & Train the denoising autoencoder model for anomaly detection. \\
    \cline{2-5}
    & \cite{kall2021asynchronous} & Code Security & Edge \& Server & Effectively and securely review and identify sensitive information in code before publishing. \\
    \cline{2-5}
    & \cite{wang2021efficient} & Fault Diagnosis & Edge \& Server & Efficiently identify possible faults in edge nodes while reducing the resource requirements and communication overhead. \\
    \cline{2-5}
    & \cite{sun2020adaptive} & IIoT & IoT \& Edge & Address the data island problem in IIoT for dynamic perception and intelligent decision. \\
    \cline{2-5}
    & \cite{lu2020communication2} & IIoT & IoT \& Edge & Improve the quality of services and implement real-time interactions in IIoT while reducing the communication cost. \\
    \cline{2-5}
    & \cite{chen2021asynchronous} & Concept Drift & Edge \& Server & Speed up model convergence for detecting and dealing with the concept drift on edge. \\
    \cline{2-5}
    \multirow{-12}{=}{\centering \rotatebox[origin=c]{90}{Smart Industry}} & \cite{sprague2018asynchronous} & Geo-location & Edge \& Server & Predict the position and orientation of the camera for end-to-end localization. \\
    \bottomrule
    \multicolumn{5}{l}{\textsuperscript{1} \footnotesize{Reference paper that belongs to the specific group.}} \\
    \end{tabular}
\end{table*}

Smart transportation is a viable situation for AFL due to its efficient utilization of computing resources that bridges the gap between training delay and time-sensitive requirements to a certain extent. 
For instance, in~\cite{lu2020blockchain}, AFL is introduced to enhance the reliability and efficiency of data sharing among vehicles. The experiments conducted in a vehicular network evaluate their scheme, including one MBS and 10 RSUs covered. The results reveal that the DAG blockchain architecture in the scheme ensures both performance and security.
Similarly, in~\cite{lu2019differentially, lu2020communication1}, AFL is adopted in urban vehicular networks to allocate resources more efficiently and securely. The experiment results verify the effectiveness of their scheme in terms of distributed data sharing and resource caching in urban vehicular networks.
In~\cite{zhang2021real}, a real-time end-to-end AFL scheme is applied in IoV and focuses on steering wheel angle prediction for autonomous driving. To conduct angle prediction, the authors utilize a two-stream deep CNN model with two separate neural branches that consume spatial and temporal information, respectively. To consume real-time streaming data, a sliding training window is introduced to reduce computation and communication latency. The experiments are carried out on real-world datasets, with the results showing that their scheme improves model prediction accuracy while reducing computation and communication latency.
AFL is also adopted in cameras to monitor, predict, and adjust traffic by controlling signal lights in the smart transportation scenario~\cite{lu2020privacy, xiaofeng2020asynchronous}. By adjusting the hyper-parameter in the scheme, an optimal balance between the model accuracy and convergence speed is achieved.
In~\cite{yang2021privacy}, AFL is adopted in unmanned aerial vehicle (UAV) networks. In order to improve the convergence speed and model accuracy, an actor-critic-based AFL scheme is proposed, including equipment selection, drone placement, resource management, local training, and global aggregation. Specifically, to prevent low-quality devices from compromising learning efficiency and model accuracy, a device selection strategy is proposed, in which nodes with high processing capability, communication capabilities, and model accuracy are selected. The selection problem is modeled as a Markov Decision Process and optimized through reinforcement learning. The scheme is evaluated by experiments, whose results show a higher learning accuracy and lower time cost.
Similarly, in~\cite{fadlullah2020hcp}, an intelligent content caching system in UAV networks based on AFL is proposed to extend the service coverage and reduce the communication delay of the 6G network. In the scheme, UAVs collaborate to forecast where content caching should be placed by taking real-time traffic distribution into account.
In~\cite{imteaj2020fedar}, AFL is applied to mobile robots that collect real-time data and perform training in a distributed and resource-constrained environment to reduce communication costs. Experiments conducted on 12 mobile robots with limited resources demonstrate that the performance of the model is guaranteed by selecting competent and reliable mobile robots.

Fault diagnosis is another application scenario for AFL.
For instance, in~\cite{ma2021asynchronous}, AFL is utilized to identify the local modes in real-time. To completely track actual system changes in real-time and increase the diagnostic rate of the nodes, each node turns private data into local models using an Extended Kalman Filter before transmitting. A sequential filter approach based on Sequential Kalman Filter is adopted to perform the asynchronous aggregation for uploaded local models. Experiments conducted on real-world collected fault datasets demonstrate high accuracy compared with benchmarks.
Similarly, in order to improve model accuracy and convergence speed in anomaly detection while preserving privacy, the authors in~\cite{tian2021towards} train the denoising autoencoder model based on labeled benign samples in AFL. Asynchronous update strategy improves the accuracy and stability of the model by reducing the impacts from the stragglers.
In~\cite{kall2021asynchronous}, the authors adopt AFL in the sensitive code review field to address privacy concerns. On leaks gathered from the code-sharing network Github, a prototype is developed and tested. When compared with local and centralized training, the proposed scheme improves model accuracy while preserving the privacy of the local training data.
Another AFL-based fault diagnosis scheme proposed in~\cite{wang2021efficient} allows nodes to adaptively select branches of the model for further training according to their local datasets. Their scheme creates an effective diagnostic model for detecting potential defects while reducing resource requirements and communication overhead. Experiments conducted across heterogeneous devices verify the feasibility of their scheme.

The AFL paradigm is also applied to IIoT environments for real-time analysis and decision-making. For example, in~\cite{sun2020adaptive}, the authors break down the barrier of data island in IIoT with the help of AFL. In their scheme, the effect of slow devices is mitigated by adaptively adjusting aggregation frequency. The experiment results validate the feasibility and efficiency of their scheme.
Similarly, in~\cite{lu2020communication2}, AFL is utilized to preserve data privacy and improve the quality of services in IIoT. Besides, by adopting digital-twin technology, real-time interactions requirements in Industry 4.0 are fulfilled. The communication cost is optimized as evidenced by experiment results.

\add{There are also some other application scenarios that adopt AFL, such as concept drift and geolocation service.} In~\cite{chen2021asynchronous}, an AFL scheme is designed to detect and handle the data distribution changes (concept drift) across edge devices. Specifically, the proposed scheme improves the predictive performance of the worst $20\%$ of devices while also maintaining the best test performance for the top $20\%$ of devices. In~\cite{sprague2018asynchronous}, the authors apply AFL to the image-based geolocation service for end-to-end localization. AFL improves the accuracy of prediction of the position and orientation of the camera while preserving the privacy of user local training data and mitigating the effects of slow devices. Experiments conducted on the CNN model across several datasets validate the feasibility.

\section{Research Challenges and Future Directions}
\label{sec:research-directions}

\add{Emerging as a trending research topic, recent studies have brought to light a collection of challenges within AFL. These challenges encompass device heterogeneity, data heterogeneity, privacy and security on heterogeneous devices, as well as applications on heterogeneous devices. To deal with these challenges, in this section, several potential research directions are identified and summarized.}

\subsection{Device Heterogeneity}

\textbf{Optimization towards balanced time cost and performance improvement: }
As summarized in section~\ref{sec:device-heterogeneity}, for AFL, the existing performance improvement strategies on heterogeneous devices, such as node selection, weighted aggregation, and cluster FL, are effective in various ways. Some of the works even adopt several strategies at the same time to improve the efficiency of AFL~\cite{chen2019communication, wu2020safa, wang2021efficient}. However, utilizing too many strategies in AFL results in a decline in efficiency to a certain extent. For instance, if selecting a range of nodes and then compressing the gradients on resource-limited devices takes longer than uploading local gradients, it is preferable to skip one of them. So far, there has been no comprehensive analysis of the balance between multiple performance improvement strategies and time consumption, which is a potential research direction. To derive the optimized trade-off, it is possible to establish a dynamic gaming model by Markov Decision Process, which can adapt to various scenarios based on the constraints. Moreover, other lightweight convex optimization methods can be considered, such as quadratic minimization with convex quadratic constraints, semidefinite programming, and convex quadratic minimization with linear constraints.

\textbf{Optimization towards generalized AFL solution: }
Usually, different performance improvement strategies have different application scenarios. For example, when the disparity in computing capabilities between heterogeneous devices is extremely high, semi-asynchronous FL with suitable weighted aggregation strategies could be an optimal solution. If the dataset distribution is IID across nodes, the local models from fast nodes should be selected and compressed, while those from slow devices should be discarded. The local models deserve higher weight if they bring a positive effect to the global model. Therefore, designing a generalized and flexible optimization framework for AFL for diverse application scenarios is a viable research field. It is expected to achieve this by integrating existing and future techniques minimally.

\textbf{Optimization towards dynamic resource allocation: }
Intuitively, AFL requires more communication resources when compared with classic FL due to more global model aggregation operations. Therefore, it is expected to consider dynamic resource allocation algorithms, including transmit power, computation frequency for model training, and model selection strategy, to maximize the long-term time average (LTA) training data size with an LTA energy consumption constraint. Specifically, a possible solution is to first define the Lyapunov drift by converting the LTA energy consumption to a queue stability constraint. Then, a Lyapunov drift-plus-penalty ratio function can be constructed to decouple the original stochastic problem into multiple deterministic optimizations along the timeline. The construction is capable of dealing with uneven durations of communication rounds. To make the one-shot deterministic optimization problem of combinatorial fractional form tractable, the fractional problem is reformulated into a subtractive-form one by the Dinkelbach method, which leads to the asymptotically optimal solution in an iterative way. By doing so, there is a potential for both higher learning accuracy and faster convergence with limited time and energy consumption.

\subsection{Data Heterogeneity}

\textbf{Optimization towards heterogeneous data distribution: }
Since the data distribution across nodes is usually non-IID in the real world~\cite{lyu2020threats}, it is meaningful to obtain a generalized model while maintaining the accuracy for each local data in AFL. There are several solutions for non-IID data challenges in classic FL, such as localized independent training~\cite{khodak2019adaptive}, personalized local model training~\cite{xu2023scei}, and cluster training~\cite{smith2017federated}. However, it is hard to transplant these solutions into AFL, since the global model prefers to convergence to nodes with higher model upload frequency (i.e. fast nodes) in AFL and result in a biased global model. Such a biased global model decreases the effects of localized independent training and personalized local model training to a certain extent. Although cluster training has been utilized in several AFL schemes, it is hard to arrange a general cluster strategy for all application scenarios. For instance, a location-based cluster strategy is not ideal for traffic prediction among smart vehicles with non-IID datasets due to the randomness of vehicle movement. Cluster training based on data distribution similarity is a potential research topic~\cite{lee2020accurate}, but it requires the development of an appropriate similarity evaluation algorithm. Besides, based on transfer learning or meta-learning, asynchronous personalized local model training is potentially an effective and accurate solution.

\textbf{Optimization towards heterogeneous data size: }
Dataset sizes among nodes are usually unequal since each node gathers its own local data independently in most AFL application scenarios. Even all nodes have identical computing resources, the imbalanced datasets across nodes lead to varying update frequencies of local models. The weighted aggregation strategy based on local dataset size is a possible solution as the work in~\cite{xiaofeng2020asynchronous}, but how to verify the validity of the dataset size on each node is another security problem. The smart contract in the blockchain offers self-verifying and self-executing capabilities~\cite{xu2021lightweight}, alleviating data fraud in AFL to some extent at a cost of low efficiency.

\textbf{Optimization towards vertical data distribution: }
Vertical data distribution is prevalent in economic scenarios, where each node possesses different feature sets of the dataset. In a heterogeneous computing environment, the lack of local models uploaded from some slow nodes causes the global model to be biased and unable to predict certain features, unlike the accuracy decline in horizontal FL. Therefore, the lagging local models are non-ignorable in vertical AFL. So far, only relatively few researches have been conducted in the vertical AFL area~\cite{xia2021vertical, chen2020vafl, gu2021privacy, zhang2021secure, li2020efficient} compared with horizontal AFL. Moreover, none of these works analyzed the effects of extreme stragglers caused by computing resources or communication resources. A possible research direction is semi-asynchronous FL. With a server-side cache, it is possible to store stale local models and increase their effectiveness while keeping other local models up to date. Besides, another potential research direction is model splitting, which splits the global model according to the feature distribution on nodes and transforms to clustered horizontal AFL, as the work in~\cite{xia2021vertical}. However, without knowing the local dataset on each node, it is hard to identify the distribution of features among nodes by the local models.

\subsection{Privacy and Security on Heterogeneous Devices}

\textbf{Optimization towards privacy protection using differential privacy: }
Differential privacy prevents AFL from a variety of privacy attacks, including background knowledge, collusion, and inference attacks. However, as the utility of local models falls, differential privacy leads global model accuracy to decline. Therefore, the trade-off between privacy and utility is hard to achieve in AFL. There are several strategies for optimizing the trade-off in classic differential privacy, such as Static Bayesian Games~\cite{qu2018improving}, Markov Decision Processes~\cite{qu2018privacy}, and Generative Adversarial Nets~\cite{qu2020gan}. However, in AFL, the publishing process for local models is dynamic and distributed, which is hard to balance through a trusted third party. Local differential privacy (LDP) is an approach that users randomly perturb their inputs without the necessity for a trusted party and is treated as a solution for the privacy issues in FL~\cite{wei2020federated}. Nevertheless, the trade-off between the privacy of local models and the utility of the global model is hard to achieve. Especially in AFL, it necessitates asynchronous macro control for the LDP in a distributed manner. The smart contract in the blockchain is a viable approach for manipulating LDP in a distributed manner. However, an asynchronous consensus algorithm needs to be designed to balance the privacy of local models and the utility of the global model.

\textbf{Optimization towards security enhancement using blockchain: }
Blockchain can be utilized to address security challenges in AFL, such as the single point of failure, Byzantine attacks, and poisoning attacks. At the same time, blockchain declines the efficiency of AFL to some extent due to its low communication efficiency and high computing resource consumption~\cite{xu2023scei}. Therefore, the trade-off between security and efficiency of blockchain-based AFL is also challenging. To improve the scalability of blockchain, several improved consensus algorithms are designed to replace PoW, such as Proof-of-Stake (PoS)~\cite{xu2021lightweight}, Proof-of-Reputation (PoR)~\cite{gai2018proof}, PBFT, and RAFT~\cite{huang2019performance}. However, generally, the higher the performance of the consensus algorithm reaches, the worse the security level is. For instance, compared with PBFT, RAFT is not resistant to Byzantine attacks but has higher data throughput. A promising solution is to develop an efficient and secure consensus algorithm. For example, Algorand~\cite{gilad2017algorand} is a byzantine-tolerant consensus algorithm with excellent scalability while maintaining a high level of security. A group of committees is randomly selected in each iteration to verify and ensure the security of the transactions. During training, the committees verify the local models in each iteration. However, it is difficult to select committees asynchronously without compromising security. In this situation, it is possible to separate the consensus process and the training process with a tailor-made blockchain structure that records the training models periodically. 

\textbf{Optimization towards security enhancement using lightweight distributed cryptography: }
Another research direction is to apply lightweight distributed cryptography to AFL to protect security. Traditional cryptography approaches, such as public-key encryption~\cite{hohenberger2005securely}, homomorphic encryption~\cite{armknecht2015guide}, and attribute-based encryption~\cite{li2020federated}, have several limitations in this case. Public-key encryption and homomorphic encryption are resource-consuming and unsuitable for resource-limited devices in AFL. Attribute-based encryption is not flexible enough, due to its necessity for a trusted third-party authority. A possible research area is to design a flexible and efficient attribute-based encryption algorithm for AFL with dynamical attribute adjustment that allows participants to join or leave freely.

\subsection{Applications on Heterogeneous Devices}

\textbf{Expansion of real-world applications: }
As summarized in section~\ref{sec:applications}, there are few real-world application scenarios for AFL for the time being, including IoV, fault diagnosis, IIoT, and so on. Compared with synchronous FL, AFL is more efficient and is better suited to time-sensitive scenarios with limited computing resources. Therefore, a possible applied research area is to apply dedicated AFL systems to a wide range of real-world scenarios. For example, in a smart hospital, ML models trained by AFL based on electronic healthcare records (EHR) predict the situation of patients. In a smart grid, AFL trains ML models on heterogeneous devices to anticipate the energy consumption in various areas and accomplish smart power dispatch. In a smart farm, the growth situation of plants is well monitored, diagnosed, and predicted by IoT with the help of ML models trained by AFL.

\textbf{Development of real-world evaluation testbeds: }
As summarized in Table~\ref{table:heterogeneous-devices}, most of the experiments of AFL are conducted in simulation mode, without demonstrating the feasibility of ALF in the real world. More experiments are expected to be conducted on IoT or edge devices to evaluate the efficiency, security, and privacy of AFL schemes. Thus, it would be a promising research direction to develop scalable and flexible testbeds deployed on heterogeneous devices and accessible from a standardized interface. The development of testbeds includes the issues of architecture design, inclusiveness of heterogeneous devices, structure-wise efficiency and performance fine-tuning, etc.

\section{Conclusion}
\label{sec:conclusion}
AFL has been attracting increasing attention due to its multiple advantageous features. To mitigate the drawbacks of existing works, three fundamental challenges in AFL are primarily studied, including device heterogeneity, data heterogeneity, as well as security and privacy issues on heterogeneous devices. By conducting an in-depth exploration of state-of-the-art research, corresponding application scenarios of AFL that potentially increase its impact and popularization on heterogeneous devices are summarized. It is pleasing to observe that the number of novel AFL schemes grows by the month. But even so, it is believed this survey is sufficiently comprehensive that new schemes can be appended and categorized correspondingly. This survey provides legible insights into the picture of AFL from a brand-new perspective, which is helpful to the community by providing potentially promising directions, and simplifying future designs, including but not limited to motivating coherent compositions uncovered by the proposed categorization and analysis.

\bibliographystyle{elsarticle-num} 
\bibliography{refer}

\end{document}